\documentclass[amsfonts,amsmath,tightenlines,preprintnumbers,nofootinbib,superscriptaddress]{revtex4}
\usepackage{graphicx}
\usepackage{epsfig}
\usepackage{bm}

\usepackage{epsfig}

\newcommand{\Choose}[2]{{\begin{pmatrix} {#1} \\ {#2} \end{pmatrix}}}

\def\pislash{ {\pi\hskip-0.6em /} }

\def\nopi{ {\rm EFT}(\pislash) }

\begin{document}

\preprint{NT@UW-07-11}
\preprint{UNH-07-02}

\begin{figure}[!t]
\vskip -1.1cm
\leftline{
{\epsfxsize=1.2in \epsfbox{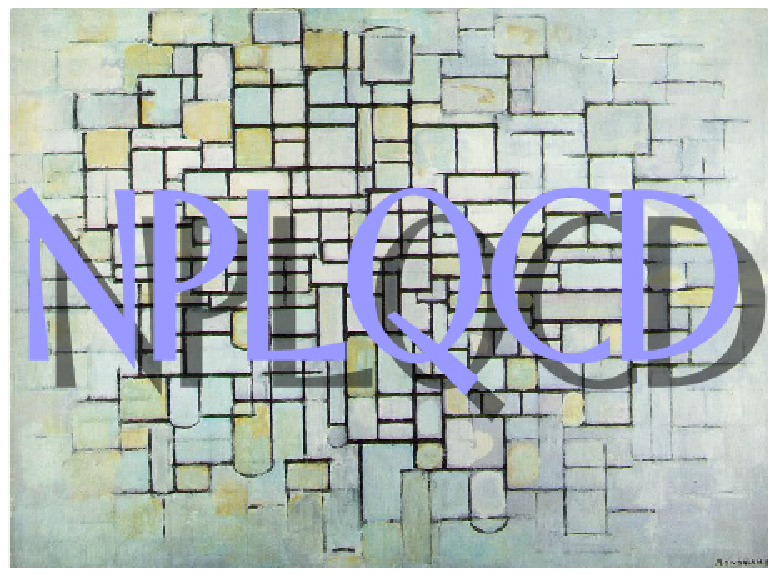}}}
\vskip 1.5cm
\end{figure}

\title{$n$-Boson Energies at Finite Volume and Three-Boson
  Interactions}

\author{Silas R. Beane} \affiliation{ Department of Physics, University
  of New Hampshire, Durham, NH 03824-3568, USA}

\author{William Detmold} \affiliation{Department of Physics,
  University of Washington, Box 351560, Seattle, WA 98195, USA}

\author{Martin J. Savage} \affiliation{Department of Physics,
  University of Washington, Box 351560, Seattle, WA 98195, USA}

\date\today

\begin{abstract}
  We calculate the volume dependence of the ground-state energy of $n$
  identical bosons with short-range repulsive interactions in a periodic spatial
  volume of side $L$, up to and including terms of order $L^{-6}$. With this result, 
  Lattice QCD calculations of the ground-state energies of three or more pions will allow for 
  a systematic extraction of the three-pion interaction at this order in the
  volume expansion.
\end{abstract}

\maketitle

\section{Introduction}
\label{sec:introduction}

A major goal of the field of strong-interaction physics is to
determine the spectrum of hadrons and nuclei from Quantum
Chromodynamics (QCD).  Lattice QCD is the only known way to rigorously
compute strong-interaction quantities, and as such, an increasing
effort is being put into understanding the lattice QCD calculations
that will be required to extract even the most basic properties of
light nuclei.  Ground-state hadron masses, including those of nuclei,
can be extracted from the long-time exponential behavior of a
two-point correlation function with the appropriate quantum numbers in
Euclidean space.  Excited states that are stable against strong
decays, such as the low-lying energy levels of nuclei, can be
extracted from the time-dependence of correlation functions over a
large time interval with various methods, such as those described in
Ref.~\cite{Michael:1985ne,Luscher:1990ck}.  By contrast, excited
states that are unstable against strong decays can be extracted by
analyzing the volume dependence of the energies of the scattering
states~\cite{Luscher:1991cf}.  However, such direct calculations of
the masses of nuclei beyond $A\sim 6$ appear computationally
prohibitive using current lattice techniques.  Consequently, the
properties of nuclei, beyond the lightest few, will likely be
systematically determined by matching lattice QCD onto nuclear
effective field theories (NEFT's)~\footnote{
For reviews of NEFT see Refs.~\cite{Beane:2000fx,Bedaque:2002mn,Epelbaum:2005pn}.
}.  Lattice QCD calculations of few
nucleon systems will be used to constrain the low-energy constants
(LECs) appearing in the NEFT, which will then be used to compute the
properties of larger nuclei. To extend calculations to larger nuclei,
further matchings (for example to the no-core shell
model~\cite{Forssen:2004dk}) will be required.  The most important
LECs relate to the interactions in the two-body sector, characterized
by the effective-range expansion of the two-particle phase
shift. These contributions (at the physical pion mass) are well
constrained by experiment, but can also be determined using lattice
QCD.  Recent work has used the volume
dependence~\cite{Hamber:1983vu,Luscher:1986pf,Luscher:1990ux} of two
nucleon~\cite{Beane:2006mx} and hyperon-nucleon~\cite{Beane:2003yx,Beane:2006gf}
systems to extract the corresponding two-body low-energy phase shifts
at specific momenta from fully-dynamical mixed-action lattice
calculations (domain-wall valence quarks on fourth-rooted
staggered configurations generated by the MILC collaboration).  In
principle, phase shifts for all momenta below inelastic thresholds
can be determined from QCD given sufficient computational resources.

In order to perform NEFT computations in systems of more than two
nucleons, $A>2$, it is necessary to include the multi-nucleon
interactions, such as the three-body force.  The leading, local
three-body interaction is a momentum-independent contact operator,
${\cal O}^{\rm 3-body}\sim D(\mu)\ [\ \overline{N}\ N\ ]^3$, where
$\mu$ is the renormalization scale (which we will discuss
subsequently), and $N$ is the nucleon annihilation operator.  A
determination of the coupling $D(\mu)$ from lattice QCD will require
the study of three-nucleon systems.  This is a complicated task, and
aside from the numerical complexities of such a lattice calculation,
theoretical developments are required in a number of relevant areas.

In this work, we compute the ground-state energy of $n$ identical bosons
in a finite cubic volume with short-range repulsive two-body
interactions, subject to periodic boundary conditions.  
This system
was first investigated for hard spheres by Huang and Yang
\cite{Huang:1957im}, and we extend their results to include order
$L^{-6}$ contributions in the volume expansion, where the two-body effective range and
three-body interactions first enter.  
For the case of three hard-spheres, our result agrees with that of
Wu~\cite{Wu:1959} (modulo renormalization issues and the appearance of a
three-body interaction).
While our results are not
directly relevant to the study of nuclei, they are necessary for
extracting the three-boson interaction in the study of multi-boson
systems (for instance multi-pion systems).

%%%%%%%%%%%%%%%%%%%%%%%%%%%%%%%%
\section{The Ground State Energy of $n$ Bosons }
\label{sec:n-boson-ground}

Motivated by interest in performing lattice QCD calculations, we
calculate the ground-state energy of $n$ bosons of mass $M$ confined
to a finite volume of size $L^3$ with periodic boundary
conditions~\footnote{The effects of the finite time extent of lattice
calculations are exponentially small in the temporal length and are
ignored herein.} in non-relativistic quantum mechanics.  In their
classic 1957 paper, Huang and Yang~\cite{Huang:1957im} (see also
Ref.~\cite{Wu:1959,Lee:1957}) considered this problem formulated for hard
spheres using the technique of pseudo-potentials. They calculated the
dependence of the $n$-boson ground state on the volume up to $L^{-5}$:
\begin{eqnarray}
  \label{eq:1}
  E_0(n, L) &=&
  \frac{4\pi\, a}{M\,L^3}\Bigg\{\Choose{n}{2}
  -\left(\frac{a}{\pi\,L}\right)\Choose{n}{2} {\cal I}
+\left(\frac{a}{\pi\,L}\right)^2 \left\{
\Choose{n}{2}{\cal I}^2
-\left[\Choose{n}{2}^2 -12\Choose{n}{3}-6\Choose{n}{4}\right]{\cal J}
\right\} \Bigg\} +{\cal O}\left(L^{-6}\right)\,,
\end{eqnarray}
where the integer sums ${\cal I}$ and ${\cal J}$ are 
\begin{eqnarray}
  \label{eq:2}
  {\cal I}&=&\lim_{\Lambda_j\to\infty}\sum_{\bf i\ne 0}^{|{\bf i}|\leq\Lambda_j} \frac{1}{|{\bf
      i}|^2} -4\pi\Lambda_j = -8.91363291781
\ \ \ ,\ \ \ 
{\cal J}\ =\ \sum_{\bf i\ne 0} \frac{1}{|{\bf
      i}|^4}  = 16.532315959
\ \ \ \ ,
\end{eqnarray}
where the sums extend over all three-vectors of integers,
and ${\tiny \Choose{n}{k}}$=$n!/(n-k)!/k!$.  In eq.~(\ref{eq:1}), the notation
has been modified, and numerical values of ${\cal I}$ and ${\cal J}$
have been corrected, compared to the expressions presented in
Refs.~\cite{Huang:1957im,Lee:1957}.  The leading term in
eq.~(\ref{eq:1}) was derived by Bogoliubov~\cite{Bog}.  In the special
case of two particles, L\"uscher~\cite{Luscher:1986pf,Luscher:1990ux}
has shown that the energy shift determined in non-relativistic quantum
mechanics, given by eq.~(\ref{eq:1}) with $n=2$, can also be derived
from quantum field theory, and thus is a general result~\footnote{ The
two-body result has been extended to the situation where the
center-of-mass is moving relative to the
volume~\cite{Rummukainen:1995vs,Kim:2005gf,Christ:2005gi}, and also to
the case in which there are coupled channels~\cite{He:2005ey}.  }.

At ${\cal O}\left(L^{-5}\right)$, the ground-state energy in
eq.~(\ref{eq:1}) is not sensitive to three-body interactions.  On
dimensional grounds, three-body contributions should first enter at
${\cal O}\left(L^{-6}\right)$ and in this work we extend the result given
above to that order.  This will allow for the extraction of three-body
interactions from lattice QCD calculations. The ground-state energy
of the $n$-boson system is calculated with an interaction of the form
\begin{eqnarray}
  \label{eq:4}
  V({\bf r}_1,\ldots,{\bf r_n}) &=&
\eta \sum_{i< j}^n \delta^{(3)}({\bf r}_i-{\bf r}_j)
+\eta_3 \sum_{i< j<k}^n \delta^{(3)}({\bf r}_i-{\bf
  r}_k)\delta^{(3)}({\bf r}_j-{\bf r}_k) +\ldots\,,
\end{eqnarray}
where the ellipsis denote higher-body interactions that do not
contribute at the order to which we work (in general, $m$-body
interactions will enter at ${\cal O}(L^{3(1-m)})$.  For an $s$-wave
scattering phase shift, $\delta(p)$, the two-body contribution to the
pseudo-potential is given by $\eta=-\frac{4\pi}{ M}
p^{-1}\tan\delta(p)= \frac{4\pi}{M}a + \frac{\pi }{M}a^2 r
(\roarrow{p}^2+\loarrow{p}^2) +\ldots$, keeping only the
contributions from the scattering length and effective range, $a$ and
$r$, respectively. At ${\cal O}(L^{-6})$ the coefficient of the
three-body potential, $\eta_3$, is momentum independent.  While up to
this point the discussion has been phrased in terms of $m$-body
pseudo-potentials, the modern language with which to describe these
interactions and calculations is that of the pionless EFT,
$\nopi$~\cite{Kaplan:1998tg,van Kolck:1998bw,Chen:1999tn}.  In $\nopi$ the divergences that
occur in loop diagrams can be renormalized order-by-order in the
expansion, preserving the power counting.  This is somewhat less
obvious when using the language of pseudo-potentials.

At ${\cal O}\left(L^{-6}\right)$ (order $V^4$, corresponding to four
insertions of the two-body potential given in eq.~(\ref{eq:4})) 
the energy-shift has contributions with interactions amongst up to six
particles. Using standard techniques of non-degenerate perturbation
theory (or the perturbative expansion of $\nopi$), 
the ground-state energy of $n$ bosons
with repulsive two-body interactions is
\begin{eqnarray}
  \label{eq:6}
  E_0(n,L) &=&
  \frac{4\pi\, a}{M\,L^3}\Bigg\{\Choose{n}{2}
  -\left(\frac{a}{\pi\,L}\right)\Choose{n}{2} {\cal I}
+\left(\frac{a}{\pi\,L}\right)^2 \left\{
\Choose{n}{2}{\cal I}^2
-\left[\Choose{n}{2}^2 -12\Choose{n}{3}-6\Choose{n}{4}\right]{\cal J}
\right\}
\\
&&\hspace*{1.5cm}+
\left(\frac{a}{\pi\,L}\right)^3\Bigg[-\Choose{n}{2} {\cal I}^3+3
\Choose{n}{2}^2 {\cal I}{\cal J} - \Choose{n}{2}^3 {\cal K} 
-24 \Choose{n}{3}\left( {\cal I} {\cal J}+2
   {\cal Q}+{\cal R}- {\cal K} \Choose{n}{2}\right) 
\nonumber
\\&&\hspace*{3cm}
-6\Choose{n}{4}\left(3 {\cal I} {\cal J}+51 {\cal K}-2
   \Choose{n}{2} {\cal K}\right) 
-300 \Choose{n}{5}{\cal K} -90
   \Choose{n}{6}{\cal K} \Bigg] \Bigg\}
\nonumber
\\&&\hspace*{3cm}
+\Choose{n}{3}\frac{64\pi a^4}{M\, L^6}(3\sqrt{3}-4\pi)\log(\mu\ L)
+\Choose{n}{2} \frac{8\pi^2 a^3}{M\, L^6}r
+\Choose{n}{3}\frac{\eta_3(\mu)}{  L^6} 
+ {\cal O}\left(L^{-7}\right)
\ \ \ \ ,
\nonumber
\end{eqnarray}
which can also be written as 
\begin{eqnarray}
\label{eq:7}
  E_0(n,L) &=&
  \frac{4\pi\, a}{M\,L^3}\Choose{n}{2}\Bigg\{1
-\left(\frac{a}{\pi\,L}\right){\cal I}
+\left(\frac{a}{\pi\,L}\right)^2\left[{\cal I}^2+(2n-5){\cal J}\right]
\\&&\hspace*{2cm}
+
\left(\frac{a}{\pi\,L}\right)^3\Big[-{\cal I}^3- (2 n-7)
  {\cal I}{\cal J} - \left(5 n^2-41 n+63\right){\cal K} -
8 (n-2) (2 {\cal Q}+{\cal R})\Big]
\Bigg\}
\nonumber
\\&&\hspace*{2cm}
+\Choose{n}{3}\frac{64\pi a^4}{M\, L^6}(3\sqrt{3}-4\pi)\log(\mu\ L)
+\Choose{n}{2} \frac{8\pi^2 a^3}{M\, L^6}r
+\Choose{n}{3}\frac{\eta_3(\mu)}{  L^6} 
+ {\cal O}\left(L^{-7}\right)
\ \ \ \ .
\nonumber 
\end{eqnarray}
In the case of $n=3$, 
the logarithmic behavior with volume was first derived
by Wu \cite{Wu:1959}, and  
the ground-state energy has recently been
computed to ${\cal O}\left(L^{-7}\right)$ by S. Tan~\cite{Tan}.  The
additional integer sums that contribute at ${\cal O}(L^{-6})$ are
\begin{eqnarray}
  \label{eq:8}
  {\cal K} &=& \sum_{\bf i\ne 0} \frac{1}{|{\bf
      i}|^6}  = 8.401923974433
\,,
\\
  \label{eq:9}
\widehat{\cal Q} &=& 
\sum_{\bf i\ne 0}\sum_{\bf j\ne 0} 
\frac{1}{|{\bf  i}|^2\ |{\bf j}|^2\ 
(|{\bf i}|^2\ +\ |{\bf j}|^2\ +\ |{\bf i}+{\bf j}|^2)}
\,,
\\
  \label{eq:10}
\widehat{\cal R} &=& 
\sum_{\bf j\ne 0}\ {1\over |{\bf j}|^4}\ 
\left[\ 
\sum_{\bf i}\ {1\over |{\bf i}|^2\ +\ |{\bf j}|^2\ +\ |{\bf i+j}|^2}
\ -\ {1\over 2}\ \int\ d^d{\bf i} \ {1\over |{\bf i}|^2}\
\ \right]
\,.
\end{eqnarray}
The sums $\widehat{\cal Q}$ and $\widehat{\cal R}$ arise from
intrinsic two-loop contributions to the energy and are naively
divergent ($\widehat{\cal R}$ contains a nested divergence, requiring
the subtraction to preserve the one-loop scattering amplitude).  As is
the case for ${\cal I}$, given in eq.~(\ref{eq:2}), we define
(regulate) these sums using dimensional regularization.  It is
straightforward to show that the dimensionally-regulated sums that
appear in eq.~(\ref{eq:9}) and eq.~(\ref{eq:10}) are
\begin{eqnarray}
  \label{eq:11}
\widehat{\cal Q}
&\to& 
{\cal Q} +\frac{4}{3}\pi^4\log(\mu\ L) - \frac{2\pi^4}{3(d-3)}
\,,
\\
  \label{eq:12}
\widehat{\cal R}
 & \to & 
{\cal R} \ -\ 2\sqrt{3}\pi^3\log(\mu\ L)\ +\  \frac{\sqrt{3}\pi^3}{d-3}
\,.
\end{eqnarray}
Numerically, we find that ${\cal Q}=-100.75569$ and ${\cal
R}=19.186903$.  In eq.~(\ref{eq:6}) and in what follows from it, the
MS-scheme is used to define the three-body coefficient, $\eta_3(\mu)$
in which only the pole, $\sim {1\over d-3}$, is subtracted from the
divergent two-loop diagrams (explicitly absorbed by $\eta_3(\mu)$).
This renormalization, and the associated logarithmic dependence on
$\mu\,L$ in eq.~(\ref{eq:6}), make clear that the three-body interaction,
$\eta_3(\mu)$, must enter at this order~\cite{Braaten:1996rq}; the
scale dependence of $\eta_3(\mu)$ exactly cancels that of the
logarithms in eq.~(\ref{eq:6}) and in subsequent expressions.  The
logarithms of $\mu\ L$ that appear in eq.~(\ref{eq:6}) result from the
relation between the momentum-space integrals in $d$-dimensions and
the integer sums.  
It is worth stressing that the quantities  ${\cal Q}$ and  ${\cal R}$
are subtraction-scheme dependent.

In the first form of our main result, eq.~(\ref{eq:6}), the
combinatoric factors make clear the number of particles involved in
the various contributions; terms proportional to ${\tiny
\Choose{n}{j}}$ have $j$ interacting particles and $n-j$ spectators.
The second, simpler form of our result, eq.~(\ref{eq:7}), follows from
evaluating the combinatoric factors.  As can be seen from these
expressions, at this order the two-body effective range also
contributes.

For two, three and four particles the ground-state energies are
\begin{eqnarray}
  \label{eq:13}
  E_0 (2,L) &=&  
\frac{4\pi\, a}{M\,L^3}\Bigg\{1
-\left(\frac{a}{\pi\,L}\right){\cal I}
+\left(\frac{a}{\pi\,L}\right)^2\left[{\cal I}^2-{\cal J}\right]
+
\left(\frac{a}{\pi\,L}\right)^3\Big[-{\cal I}^3 +3
  {\cal I}{\cal J} - {\cal K}\Big]
\Bigg\}+\frac{8\pi^2 a^3}{M\,L^6}r
 + {\cal O}\left(L^{-7}\right)
\ ,
\end{eqnarray}
\begin{eqnarray}
  \label{eq:13b}
  E_0 (3,L) &=&   \frac{12\pi\, a}{M\,L^3}\Bigg\{1
-\left(\frac{a}{\pi\,L}\right){\cal I}
+\left(\frac{a}{\pi\,L}\right)^2\left[{\cal I}^2+{\cal J}\right]
%\\&&\hspace*{2cm}
+
\left(\frac{a}{\pi\,L}\right)^3\Big[-{\cal I}^3 + 
  {\cal I}{\cal J} + 15 {\cal K} -
8 (2 {\cal Q}+{\cal R})\Big]
\Bigg\}
\\&&\hspace*{2cm}
+\frac{64\pi a^4}{M\, L^6}(3\sqrt{3}-4\pi)\log(\mu\ L)
+\frac{24\pi^2 a^3}{M\, L^6}r
+\frac{1}{L^6} \eta_3(\mu)
 + {\cal O}\left(L^{-7}\right)
\ \ \ ,
\nonumber
\end{eqnarray}
\begin{eqnarray}
  \label{eq:13c}
  E_0 (4,L) &=&   \frac{24\pi\, a}{M\,L^3}\Bigg\{1
-\left(\frac{a}{\pi\,L}\right){\cal I}
+\left(\frac{a}{\pi\,L}\right)^2\left[{\cal I}^2+3{\cal J}\right]
%\\&&\hspace*{2cm}
+
\left(\frac{a}{\pi\,L}\right)^3\Big[-{\cal I}^3 -
  {\cal I}{\cal J} + 21 {\cal K} -
16 (2 {\cal Q}+{\cal R})\Big]
\Bigg\}
\\&&\hspace*{2cm}
+\frac{256\pi a^4}{M\, L^6}(3\sqrt{3}-4\pi)\log(\mu\ L)
+\frac{48\pi^2 a^3}{M\, L^6}r
+\frac{4}{L^6} \eta_3(\mu)
 + {\cal O}\left(L^{-7}\right)\,.
\nonumber
\end{eqnarray}
From eq.(\ref{eq:6}), it is possible to construct combinations of
these energies that are directly sensitive to the three-body
contributions at $L^{-6}$. That is, for $n>2$, 
\begin{eqnarray}
  \label{eq:16}
&& E_0(n,L) - \Choose{n}{2}  E_0(2,L) 
\nonumber\\
&&\qquad\qquad
\ -\  
6 \Choose{n}{3} \left[\ 
\frac{M^2 L^4 }{16 \pi ^4}\ {\cal J} \left[ E_0(2,L) \right]^3 
\ +\ \frac{M^3 L^6}{32 \pi ^6}\ 
 \left({\cal I} {\cal J}-\frac{1}{4} {\cal K} (5
   n-31)-2 (2 {\cal Q}+{\cal R})\right) \left[ E_0(2,L) \right]^4\ \right]
\nonumber\\
&& \qquad\qquad\qquad\qquad
\ =\ 
\Choose{n}{3} 
\frac{1}{L^6}\left[\frac{64\pi a^4}{M}(3\sqrt{3}-4\pi)\log(\mu\ L)
+ \eta_3(\mu)\right] 
 + {\cal O}\left(L^{-7}\right)
\ \ \ .
\end{eqnarray}
This can also be accomplished by forming judicious combinations of the energies of
three different systems, such as $E_0(4,L), E_0(3,L)$ and $E_0(2,L)$.
Further, combinations of energies can be formed in which the three-body
interaction is absent, such as 
\begin{eqnarray}
  \label{eq:no3}
E_0(2,L) - {2\over 3}  E_0(3,L) + {1\over 6} E_0(4,L) 
\ +\  40 \ {M^3 L^6\over 256\pi^6} \ {\cal K}\  \left[ E_0(2,L) \right]^4
& = &  {\cal O}\left(L^{-7}\right)
\ \ \ .
\end{eqnarray}
Relations such as that of eq.~(\ref{eq:no3}) will provide  a useful check of 
both the statistical and systematic uncertainties associated with 
lattice QCD calculations.

The calculation of ground-state energies described here has been
derived in a non-relativistic framework, however the results remain
valid relativistically.  In the two-body case, this has been shown by
L\"uscher~\cite{Luscher:1986pf}.  In the higher-body case, the
non-relativistic calculation will not correctly recover a
field-theoretic calculation, due to relativistic effects in multiple,
two-body interactions involving three or more particles.  At ${\cal
O}\left(L^{-4}\right)$, only two-particle interactions contribute to
the $n$-body ground-state energy and the results of
Ref.~\cite{Luscher:1986pf} follow without modification.  Since the
interaction of three particles due to the two-body interaction first
enters at $L^{-5}$, and relativistic effects in such interactions are
suppressed by $(ML)^{-2}$, the first relativistic effects will occur
at ${\cal O}(L^{-7})$.  Generally, relativistic effects can be
included perturbatively into the volume expansion by using
$\nopi$~\cite{Kaplan:1998tg,van Kolck:1998bw,Chen:1999tn}.

Beyond ${\cal O}(L^{-6})$ a number of other effects are also important:
\begin{itemize}
    \item 
Higher partial waves will contribute to the ground-state energy at
finite volume.  The cubic periodic boundary conditions imposed on the
system are such that the spatial symmetry group is $H(3)$ with the
ground-state wavefunction transforming in the $A_1^+$ representation
(see Ref.~\cite{Mandula:1983ut}).  Contributions to the ground-state
energy from interactions other than $s$-wave (as classified in the
continuum) first enter at order ${\cal O}(L^{-7})$ via three two-body
interactions, two of which are $s$-wave and one of which is $p$-wave.
The leading two-body contribution from higher partial-waves results
from a single insertion of an $l=4$ operator, and enters at ${\cal
O}(L^{-11})$.
\item 
It is straightforward to show that the leading, momentum-independent
$m$-body operators contribute to the ground-state energy at
$L^{3(1-m)}$, and therefore, the four-body local operator will first
enter at order $L^{-9}$.  It is obvious that the momentum-dependent
$m$-body operators first enter two orders higher than the
momentum-independent operators.  For instance, contributions from the
two-body shape parameters (higher-order terms in the expansion of
$p\cot\delta(p)$) first occur at ${\cal O}(L^{-8})$.

\item 
In the calculations that we have performed, the symmetrizations that
are required to satisfy Bose statistics are trivial.  At higher
orders in the expansion, where multiple insertions of the two-body
effective-range interaction occur, symmetrization of the intermediate
states will have non-trivial consequences and calculations will become
increasingly complicated.
\end{itemize}

%%%%%%%%%%%%%%%%%%%%%%%%%%%%%%%%%%
\section{Discussion and Conclusion}
\label{sec:discussion}

In this work we have computed the volume dependence of the
ground-state energy of $n$ bosons in a cubic volume with periodic
boundary conditions.  Knowledge of this dependence is necessary in
order to establish a connection between lattice QCD calculations of
$n$-boson systems in Euclidean space and the multi-body interactions
contributing to the properties of many-body systems in Minkowski
space.  We conclude by highlighting a number of issues and possible
extensions of this work:
\begin{itemize}
\item
Our results are easily generalized to asymmetric volumes. In the
asymmetric case, the integer sums that appear in the expressions for
the ground-state energy will become dependent on the asymmetry
parameters~\cite{Li:2003jn,Detmold:2004qn} (for some asymmetries, one
or more of these sums can vanish~\cite{Detmold:2004qn}).  Further, the
reduced symmetry of the system forces derivative interactions and
higher partial-wave interactions to enter at lower orders in the
volume expansion~\cite{Feng:2004ua}.
\item
Calculation of excited-state energies are straightforward and tedious.
Many contributions (diagrams in the perturbative expansion)
that vanish in the calculation of the ground-state
energy will no longer vanish, and the combinatoric factors become
somewhat more complicated because of the symmetrization requirements.
\item
Our calculation of the
ground-state energy of the $n$-boson system is an expansion whose
terms  depend parametrically on $n$ and $L$ as
$n^\alpha/L^\beta$, where $\alpha$ and $\beta$ are positive integers.
Much of the discussion in the literature regarding collections of
bosons concerns Bose-condensed systems, with the 
results that
are summarized in Ref.~\cite{Braaten:2000eh}.  
It is clear that in the large-$n$ limit, relevant to the
Bose-condensed systems, this expansion fails, and as such, no direct
connection between our calculation and Bose-condensed systems can be
made.
\item
We have restricted ourselves to the case of repulsive interactions
among the bosons.  The reason for this is that for an attractive
two-body interaction of sufficient strength, the ground-state will not
be perturbatively close to $n$ non-interacting bosons but will, in
fact, be a system of two-body (or higher body) bound states, e.g. one
can imagine that for the three-body system the ground-state is a
two-body bound state interacting with the third boson.  The
finite-volume behavior of the two-particle system with shallow
(compared to the inverse range of the underlying interaction) bound
states was discussed in Ref.~\cite{Beane:2003da}, and subsequently
numerically explored in Ref.~\cite{Sasaki:2006jn}. 
In such systems, the leading finite volume corrections have both power-law and
exponential volume dependences, and are somewhat
more complicated to analyze.
\item
The expressions we have presented in this work form the basis for
investigating $\pi^n$ correlation functions with isospin $I = I_z = n$
in lattice QCD.  By forming well-defined ratios of correlation
functions, as given in eq.~(\ref{eq:16}), the three-body interaction
can be calculated.  The natural choice of renormalization scale is the
ultra-violet cutoff of the low-energy theory, which is
approximately the range of the interaction, $\mu\sim r^{-1}$.  This
gives rise to a large logarithm, $\log(r L)/L^6$, which should
formally dominate the $L^{-6}$ contribution to the ground-state
energy.
\item
As the lattice-QCD study of nuclei is the underlying motivation for
this work, it is worth considering difficulties that will be
encountered in generalizing the result described here to fermionic
systems.  A significant difference between the three-neutron system
and the three-boson system is that the Pauli-exclusion principle
forces the unperturbed ground state of three neutrons to have non-zero
momentum.  One of the neutrons will have momentum $|{\bf p}|=2\pi/L$,
which provides power-law volume dependence to the ground state even in
the absence of interactions. As a result the momentum-dependent
interactions will enter at lower orders than in the boson system.  The
lowest-energy state with zero total momentum consists of one neutron
at rest, and the other two moving back-to-back, each with $|{\bf
p}|=2\pi/L$.
\end{itemize}

\acknowledgements{We thank A. Bulgac, M.~Forbes, S. Tan and A. Torok   for useful
  discussions. The work of SRB was supported in part by National
  Science Foundation CAREER grant No. PHY-0645570 and that of WD and
  MJS by Department of Energy grant DE-FG03/974014. }

\end{document}